\UseRawInputEncoding
\documentclass[pra,twocolumn,showpacs,floatfix]{revtex4}
\usepackage{graphicx}
\usepackage{color}
\usepackage{amsmath}
\begin{document}

\title{Many-body state and dynamic behaviour of the pair-correlation function of a small Bose-Einstein 
condensate confined in a ring potential}

\author{A. Roussou$^{1,2}$, J. Smyrnakis$^1$, M. Magiropoulos$^1$, and G. M. Kavoulakis$^{1,3}$}

\affiliation{$^1$Hellenic Mediterranean University, P.O. Box 1939, GR-71004, Heraklion, Greece \\
$^2$Department of Physics, Chalmers University of Technology, SE-412 96, G\"oteborg, Sweden \\
$^3$HMU Research Center, Institute of Emerging Technologies, P.O. Box 1939, GR-71004, Heraklion, Greece}

\date{\today{}}

\begin{abstract}

We investigate the many-body state and the static and the dynamic behaviour of the pair-correlation 
function of a Bose-Einstein condensate with a finite atom number, which is confined in a quasi-one-dimensional 
toroidal/annular potential, both for repulsive, and for attractive interactions. We link the dynamic 
pair-correlation function that we evaluate with the problem of quantum time crystals. For weak repulsive 
interatomic interactions and a finite number of atoms the pair-correlation function shows a periodic temporal 
behaviour, which disappears in the limit of a large atom number, in agreement with general arguments. Finally
we provide some insight into older results of attractive interactions, where the time-crystalline behaviour 
exists only in the limit of a large atom number.  

\end{abstract}
\pacs{05.30.Jp, 03.75.−b, 03.75.Kk} \maketitle

\section{Introduction}

One problem which has attracted attention in recent years is the realization of ``quantum time 
crystals". Initially this concept was introduced by Wilczek \cite{FW1}, and Shapere and Wilczek 
\cite{FW11} and it refers to a system which is stable and conservative, breaking continuous time 
translation symmetry. In simple terms, the question is whether a physical system may realize a 
``perfect" clock, which is conservative and shows a periodic behaviour. 

Soon after Refs.\,\cite{FW1, FW11} were published, a series of papers followed \cite{Brunocom, 
Wilcom, Bruno2, Bruno3, Nozieres, Watanabe, KS, Laz, Else, Heis, Yao, Sachaprl, Nay1, PO, Watanabe2,
Sacha2}. These included also the case of driven systems, where the concept of discrete time crystals 
was introduced. On the experimental side, time crystals have been realized in a disordered dipolar 
many-body system \cite{Choi} and in a spin chain of trapped atomic ions \cite{Zhang}. Numerous 
other papers have been published, discussing this interesting question. The literature on this 
problem is rather rich and here we just refer to some recent review articles, which give an 
overview of the work that has been done on this field so far \cite{Sacha, Sondhi, Guo, Nayak}.

The problem of quantum time crystals has turned out to be rather controversial. It resembles the 
one of spontaneous breaking of space translation symmetry. Bruno \cite{Bruno2, Bruno3} considered a 
many-body system in an Aharonov-Bohm ring under the action of a potential that rotates periodically 
along the ring. He showed that for a ground state that breaks rotational symmetry, the moment of 
inertia of the system is always positive and imposing rotation increases the energy. Watanabe and 
Oshikawa \cite{Watanabe} avoided the use of a symmetry-breaking perturbation and studied correlation 
functions. They showed that, in the ground state of the system and in the thermodynamic limit of an 
infinite system the two-point correlation function is time-independent.

An ideal system for studying superfluid and many-body effects is that of Bose-Einstein condensed 
atoms, which are confined in a ring potential. In the present study we consider this system, assuming 
repulsive, and attractive interatomic interactions, and examine its many-body state and its connection 
with the problem of time crystals. As we analyse in detail below, the case of repulsive interactions 
is very different compared with the case of attractive interactions. 

When one is working with the diagonalization of the many-body Hamiltonian, it is well-known that the 
single-particle density distribution is always axially symmetric, due to the assumed axial symmetry 
of the Hamiltonian. One way to break this symmetry is via the pair-correlation function, as in 
Refs.\,\cite{Sachaprl, Sacha2}, where, as mentioned also above, attractive interactions were considered. 
The pair-correlation function may be viewed as the probability of observing an atom at a specific point, 
under the condition that another atom is fixed at a different point. Alternatively, this quantity gives 
the single-particle density distribution at a certain point, after the removal of a single atom at 
another point \cite{pco}.

According to the results that follow below, for repulsive interactions, the derived pair-correlation 
function in the ground state of the system that we consider shows a periodic temporal behaviour, 
only for a finite atom number. While it is tempting to think of this system as a time crystal, still 
this is not the case, since one has to consider the thermodynamic limit of an infinite system 
\cite{Bruno2, Bruno3, Watanabe}. Actually, the fact that in this limit the pair correlation function 
that we evaluate tends to a constant, is in agreement with the general arguments of Watanabe and 
Oshikawa \cite{Watanabe}. The case of attractive interactions is very different. When the interactions
are sufficiently strong, the cloud forms a localized blob \cite{Carr}; see also \cite{Ueda1, GK, Ueda2}.
Actually, this system has already been considered in the problem of time crystals, for effectively 
attractive interatomic interactions \cite{FW1, Sachaprl, Sacha2}. While in the limit of a large atom
number we do get a time crystal (provided that the localized blob rotates around the ring, performing
solid-body-like rotation), for a finite atom number the density profile changes and the time-crystalline
behaviour is lost \cite{Sacha2}. 

Various reasons make this study timely and interesting, in addition to the insight that it provides
into the more general problem of time crystals. First of all, it is closely related with the so-called 
field of ``atomtronics" \cite{at1, at2}. Secondly, numerous experiments on cold atoms have been performed 
in toroidal/annular traps, see, e.g., Refs.\,\cite{Sauer, Kurn, Arnold, Olson, Phillips1, Heathcote, 
Henderson, Foot, GKK, Moulder, Zoran, Ryu, WVK, hysteresis, hyst2, Perin, WVK2}. Furthermore, it examines 
the many-body state of this system in the case of a finite number of atoms; we stress that there is a 
tendency in this field towards the study of systems with a finite and even small atom number \cite{SJ}. 
Finally, it presents the dynamic behaviour of the pair correlation function of the well-known Lieb-Liniger 
model \cite{LLM} (which we adopt in this study).  

As we see below, there are two hierarchies in this system. The first is the one associated with 
different powers of the atom number. The second comes from the amplitudes of the basis states of 
the many-body state, which drop exponentially for weak interactions. Taking advantage of these 
two hierarchies, in addition to the numerical results that we derive, we also present analytic 
results, which shed light into this quantum system and its temporal behaviour. 

In what follows below we present in Sec.\,II our model. In Sec.\,III we examine the many-body 
state for both repulsive and attractive interactions. In Sec.\,IV we first evaluate the time-independent 
pair correlation function, and then we turn to the time-dependent problem. Finally, in Sec.\,V 
we summarize our results, we present our conclusions and we comment on the experimental relevance 
of our study. 
 
\section{Model}

Let us thus turn to our model. As mentioned above, the actual problem we have in mind is that of an 
annular/toroidal trap, which is very tight in the transverse direction. As long as the quantum of 
energy in the transverse direction is much larger than the interaction energy, the atoms reside in 
the lowest mode of the potential (in the transverse direction) and the system effectively becomes 
one-dimensional \cite{JKP}. Therefore, the Hamiltonian that we consider is essentially the Lieb-Liniger 
model \cite{LLM}. Setting $\hbar = 2 M = R = 1$, where $M$ is the atom mass, and $R$ is the radius of 
the ring, is
\begin{equation}
  H = \sum_n n^2 c_n^{\dagger} c_n 
  + \frac g {2 \pi} \frac 1 2 \sum_{m, n, k, l} c_m^{\dagger} c_n^{\dagger} c_k c_l \, \delta_{m+n, k+l},
  \label{Hamm}
\end{equation}
where $g$ is the (effective) matrix element for s-wave, elastic atom-atom collisions. Also, $c_k$ is the 
annihilation operator of a particle with angular momentum $k$, being in the eigenstate of the ring 
potential $\psi_k = e^{i k \theta}/\sqrt{2 \pi}$, with $\theta$ being the angle. The corresponding 
eigenenergy is given by $\epsilon_k = k^2$.

The eigenstates of the Hamiltonian are also eigenstates of the angular momentum, and as a result the 
single-particle density 
\begin{eqnarray}
n(\theta) = \langle \Psi_1 (N, L = 0)| \Phi^{\dagger}(\theta) \Phi(\theta) | \Psi_1 (N, L = 0) \rangle, 
\end{eqnarray}
is equal to $N/(2 \pi)$, i.e., axially symmetric, with $N$ being the atom number. Here $\Phi(\theta)$ 
is the operator which destroys an atom at $\theta$, with 
\begin{eqnarray}
\Phi(\theta) = \sum_k c_k \psi_k(\theta) = \sum_k c_k e^{i k \theta}/\sqrt{2 \pi}. 
\end{eqnarray}
Also, $|\Psi_p (N, L) \rangle$ denotes the $p$th excited eigenstate (with $p=1$ being the ground state) 
of the many-body Hamiltonian with $N$ atoms and $L$ units of angular momentum, with a corresponding 
eigenenergy ${\cal E}_p(N,L)$.
 
\section{Many-body state}

\subsection{Repulsive interactions}

Let us consider the limit of weak and repulsive interactions, $\gamma = N g/(2 \pi) \ll \epsilon_1 = 1$, 
which allows us to work in the truncated space that includes the single-particle states $\psi_0$ and 
$\psi_{\pm 1}$ only. The many-body ground state of the system with $N$ atoms and $L=0$ units of total 
angular momentum may be expressed as 
\begin{eqnarray}
  |\Psi_1(N, L=0) \rangle = \sum_m (-1)^m d_m |(-1)^m, 0^{N - 2m}, (+1)^m \rangle.
  \label{fock}
\end{eqnarray}
The notation $|(-1)^m, 0^{N - 2m}, (+1)^m \rangle$ means that there are $m$ atoms in the states 
$\psi_{\pm 1}(\theta)$ and $N-2m$ atoms in the state $\psi_0(\theta)$. Also, in the above expression 
$d_m$ are positive, while the term $(-1)^m$ comes from the minimization of the energy. The Hamiltonian 
may be diagonalized using the Bogoliubov transformation \cite{Ueda1, GK, Ueda2}. Since we are interested 
in the amplitudes $d_m$, let us write the eigenvalue equation, which has the form \cite{and}
\begin{eqnarray}
  - H_{m, m-1} d_{m-1} + H_{m, m} d_m - H_{m, m+1} d_{m+1} = {\cal E}_1 d_m,
\end{eqnarray}
where $H_{m,n} = \langle m | H | n \rangle$ are the matrix elements of the Hamiltonian. For the diagonal ones,
\begin{eqnarray}
H_{m,m} = \gamma (N-1)/2 + 2m (1+\gamma), 
\end{eqnarray}
while the off-diagonal, 
\begin{eqnarray}
H_{m, m+1} \approx \gamma (m+1), 
\end{eqnarray}
for $m \ll N$ \cite{GK}. From the above equation and for the approximate expressions of the matrix elements, it 
follows that, for small $\gamma$, the amplitudes 
\begin{eqnarray}
 d_m \approx (\gamma/2)^m
 \label{amplt}
\end{eqnarray}  
for $m \ge 1$, and also 
\begin{eqnarray}
{\cal E}_1 \approx \gamma (N-1)/2 - \gamma^2/2. 
\end{eqnarray}
We stress that in the many-body state of Eq.\,(\ref{fock}), the density matrix is diagonal, with its 
eigenvalues being the occupancy of the three single-particle states. We find that
\begin{eqnarray}
  \langle c_{\pm 1}^{\dagger} c_{\pm 1} \rangle = \sum_m m d_m^2 \approx \left( \frac {\gamma} 2 \right)^2,
\end{eqnarray}
while 
\begin{eqnarray}
  \langle c_{0}^{\dagger} c_{0} \rangle = \sum_m (N-2m) d_m^2 \approx N - 2 \left( \frac {\gamma} 2 \right)^2.
\end{eqnarray}
Therefore, only one of these three eigenvalues scales linearly with $N$, while the other two are independent 
of $N$, for fixed $\gamma$, as expected for a (non-fragmented) Bose-Einstein condensed system. In the limit
of large $N$, the many-body state becomes the trivial state
\begin{eqnarray}
  |\Psi_1(N, L=0) \rangle \approx |(-1)^{0}, 0^{N}, (+1)^{0} \rangle,
  \label{fockk2}
\end{eqnarray}
i.e., it reduces to the trivial mean-field, product, state, of the form  
\begin{eqnarray}
\Psi_{\rm MF}(\theta_1, \theta_2, \dots, \theta_N) = \prod_{i=1}^N \psi_{0}(\theta_i).
\label{mfrep}
\end{eqnarray}

In addition to the above analytic results, we have diagonalized numerically the Hamiltonian of Eq.\,(\ref{Hamm})
within some set of single-particle orbitals $\psi_q$, with $q_{\rm min} \le q \le q_{\rm max}$, that we can tune.
More specifically, we construct the Fock states with some given atom number $N$ and angular momentum $L$ (clearly
$L = 0$ in this case) and diagonalize the resulting many-body Hamiltonian of Eq.\,(\ref{Hamm}). Figure 1 shows the 
five largest amplitudes $d_m$ that result from such a calculation in the truncated space of Eq.\,(\ref{fock}), for 
$N = 100$ atoms, with $\gamma = 0.05$. In the same plot we also show the analytic expression of Eq.\,(\ref{amplt}). 
The difference between the two sets of data is hardly visible.

\begin{figure}[t]
\includegraphics[width=9.5cm,height=4.cm,angle=-0]{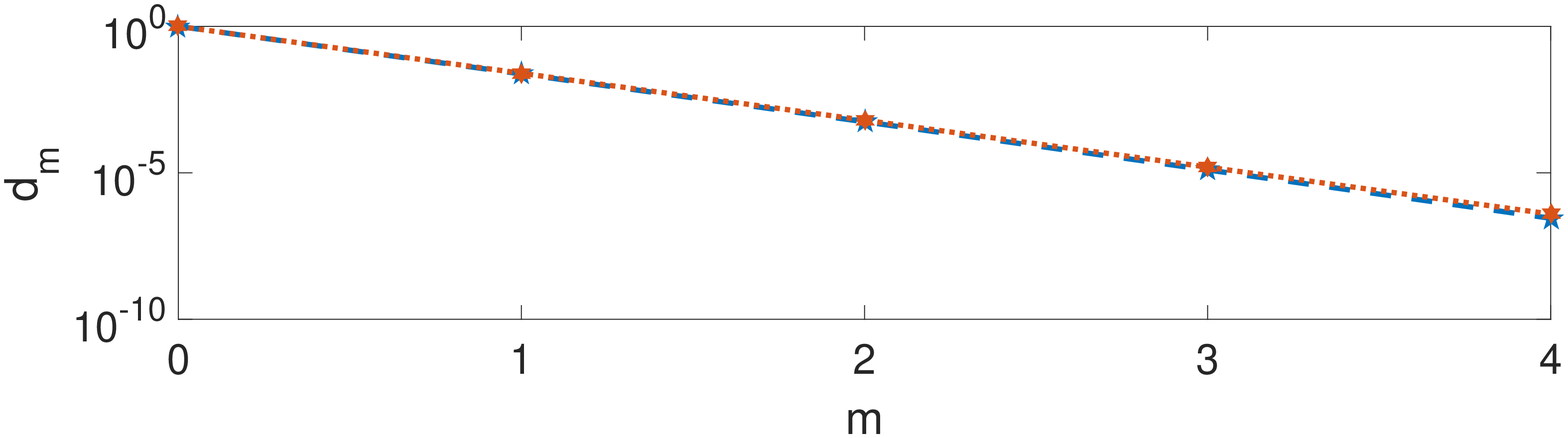}
\caption{(Color online) The five largest amplitudes $d_m$ of Eq.\,(\ref{fock}), evaluated numerically from 
the diagonalization of the many-body Hamiltonian (blue, dashed curve) and the analytic expression $d_m = 
(\gamma/2)^m$ (orange, dotted curve), for $N = 100$ atoms and $\gamma = 0.05$.} 
\end{figure}

\begin{figure}[t]
\includegraphics[width=9.5cm,height=4.cm,angle=-0]{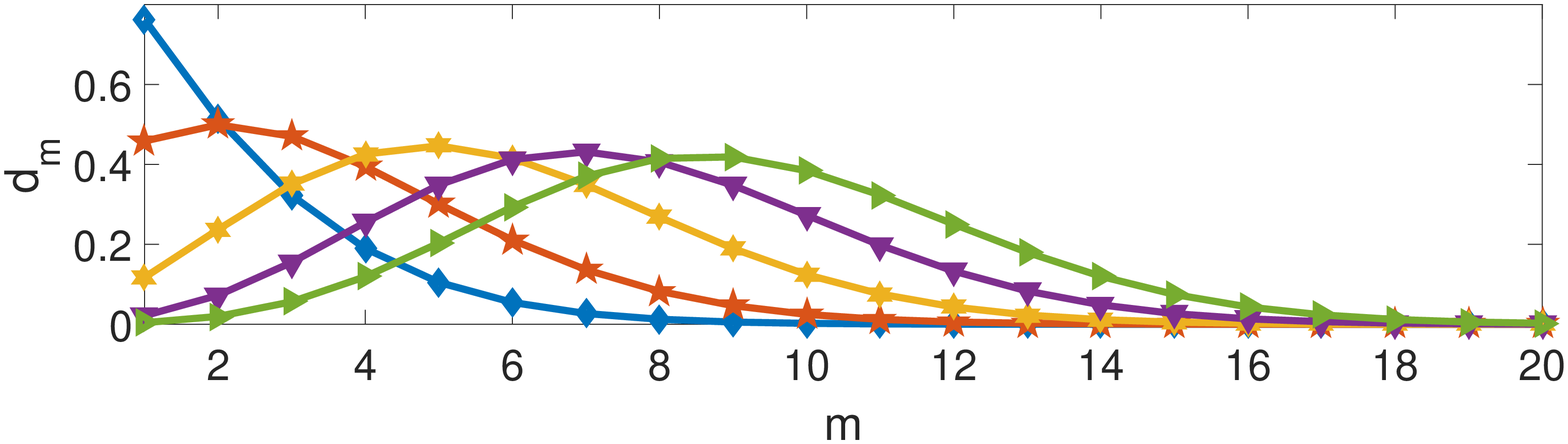}
\caption{(Color online) The amplitudes $d_m$ of Eq.\,(\ref{fock}), evaluated numerically from the diagonalization 
of the many-body Hamiltonian, for $N = 100$ atoms and $\gamma = -0.5, -0.55, -0.6, -0.65$, and $-0.7$, from the
lowest to the highest one, on the right side of the plot.} 
\end{figure}

\begin{figure}[h]
\includegraphics[width=9.5cm,height=4.cm,angle=-0]{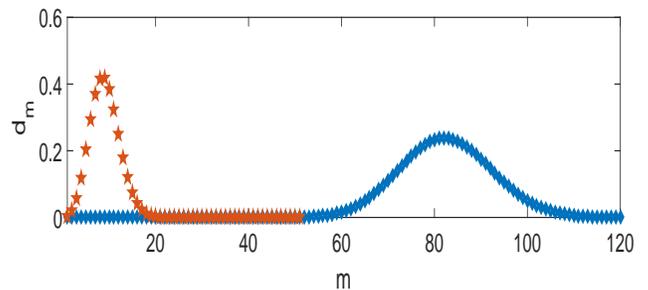}
\caption{(Color online) The amplitudes $d_m$ of Eq.\,(\ref{fock}), evaluated numerically from the diagonalization 
of the many-body Hamiltonian for $N = 100$ (left) and $N = 1000$ atoms (right), and $\gamma = -0.7$.} 
\end{figure}

\subsection{Attractive interactions}

At this point it is worth examining the case of attractive interactions, where, as we mentioned also above, 
for a sufficiently strong attractive interaction strength, the cloud forms a localized blob \cite{Carr, Ueda1, 
GK, Ueda2}. In this case where the effective interaction is attractive, minimization of the energy implies that 
the phase $(-1)^m$ in Eq.\,(\ref{fock}) is absent,
\begin{eqnarray}
  |\Psi_1(N, L=0) \rangle = \sum_m d_m |(-1)^m, 0^{N - 2m}, (+1)^m \rangle.
  \label{fockkk}
\end{eqnarray}
In Fig.\,2 we consider $N = 100$ atoms and various values of $\gamma = -0.5, -0.55, -0.6, -0.65$ and $-0.7$.
Within the mean-field approximation the critical value for the transition from a homogeneous state to a localized 
is $\gamma = -1/2$ \cite{Ueda1, GK}. In Fig.\,2 we see clearly this transition, where, for this small system, the
critical value of $\gamma$ is shifted due to the finiteness of $N$ \cite{Ueda1, GK}. We observe that as $\gamma$ 
becomes more negative, the amplitudes develop a non-monotonic behaviour, which is associated with the fact that 
all three eigenvalues of the density matrix scale linearly with $N$ and the system becomes fragmented. As a result, 
the nature of the problem changes completely, as compared to the case of repulsive interactions. In more physical 
terms, the single-particle density distribution (within the mean-field approximation) becomes inhomogeneous.

Furthermore, in Fig.\,3 we have plotted the amplitudes for $N = 100$ and $N = 1000$ atoms, and a fixed $\gamma = 
-0.7$. Here we see that the value of $m$ where the maximum of the amplitudes occurs scales (roughly) linearly with 
$N$, while the width is of order $\sqrt{N}$. As a result, in the limit of large $N$, with $\gamma$ fixed and smaller 
than $-1/2$, the many-body state becomes, in this case, 
\begin{eqnarray}
  |\Psi_1(N, L=0) \rangle \approx |(-1)^{N_{-1}}, 0^{N_0}, (+1)^{N_1} \rangle,
  \label{fockk}
\end{eqnarray}
with all three $N_i$ being of order $N$, and thus it reduces to the mean-field, product, state 
\begin{eqnarray}
\Psi_{\rm MF}(\theta_1, \theta_2, \dots, \theta_N) = 
\nonumber \\
 = \prod_{i=1}^N [a_{-1} \psi_{-1}(\theta_i) + a_{0} \psi_{0}(\theta_i) + a_{1} \psi_{1}(\theta_i)],
\label{mfatt}
\end{eqnarray}
with $|a_{-1}|^2 = N_{-1}/N, |a_0|^2 = N_0/N$, and $|a_1|^2 = N_1/N$ being the occupancy of the single-particle states 
$\psi_{-1}, \psi_0$, and $\psi_1$, respectively.

\section{Pair-correlation function}

\subsection{Time-independent problem}

As we mentioned above, the single-particle density distribution $n(\theta)$ is always axially symmetric. Therefore, 
the density is not a helpful observable. We thus turn to the pair-correlation function \cite{Sachaprl}, which is 
defined as 
\begin{eqnarray}
  n^{(2)}(\theta, \theta') = 
  \frac {\langle \Psi_1 | \Phi^{\dagger}(\theta') \Phi^{\dagger}(\theta) \Phi(\theta) \Phi(\theta') | \Psi_1 \rangle}
  {\langle \Psi_1 | \Phi^{\dagger}(\theta)  \Phi(\theta) | \Psi_1 \rangle 
   \langle \Psi_1 | \Phi^{\dagger}(\theta') \Phi(\theta')| \Psi_1 \rangle}.
\nonumber \\
   \label{corr}
\end{eqnarray}
Because of the axial symmetry of our problem each term in the denominator, which is the single-particle 
density distribution, is a constant. For the same reason, $n^{(2)}(\theta, \theta')$ is a function of the 
difference $\Delta \theta = \theta - \theta'$. Finally, we stress that, very generally, $g N^2 n^{(2)}(\theta, 
\theta)/(4 \pi)$ is equal to the expectation value of the interaction energy, which follows
directly from the interaction term of the Hamiltonian of Eq.\,(\ref{Hamm}). 
 
In what follows in the rest of the paper we are mostly concerned about the case of repulsive interactions. 
Still, we also comment briefly on the case of attractive interactions at the end of this section. Returning 
to Eq.\,(\ref{corr}), and given the results of Sec.\,III A, within the truncated space that we have considered, 
there are three classes of terms. First of all, we have the term $(c_0^{\dagger})^2 c_0^2$, which is of order 
$N^2$. The second class of terms includes the ones with two operators having index $``0"$, and the other two 
$``+1"$ and/or $``-1"$, which are of order $N$. In the third class of terms we have the operators with index 
$``+1"$ and/or $``-1"$, only, which are of order unity.

The additional hierarchy of terms that plays a crucial role, especially in the dynamics that is described 
below, is associated with the rapid -- exponential -- decay of the amplitudes $d_m$ in the many-body (ground) 
state, as discussed above. Clearly $d_0$ is the dominant one, being of order unity, $d_0 = 1 + {\cal O}(d_1^2)$. 
For small values of $\gamma$, which may serve as our ``small" parameter, very few of the amplitudes $d_m$ are 
non-negligible. In this limit, one may thus assume that the ground state $|\Psi_1 (N, L=0) \rangle$ is given 
by the first two terms only in Eq.\,(\ref{fock}), 
\begin{eqnarray}
|\Psi_1(N, L=0) \rangle &\approx& d_0 |(-1)^0, 0^N, (+1)^0 \rangle 
\nonumber \\
&-& d_1 |(-1)^1, 0^{N-2}, (+1)^1 \rangle.
 \label{2st}
\end{eqnarray} 
Then,
\begin{eqnarray}
  n^{(2)}(\theta, \theta') = \frac {N-1} N - 4 d_0 d_1 \frac {\sqrt{N (N-1)}} {N^2} \cos(\Delta \theta) + 
  \nonumber \\
 + \frac {d_1^2} {N^2} [4 (N-2) \cos(\Delta \theta) + 2 \cos(2 \Delta \theta)],
 \label{ws}
\end{eqnarray}
where $\Delta \theta = \theta' - \theta$. As we see, this quantity is spatially dependent, however the 
spatial dependence is an effect of the finiteness of $N$, which becomes negligible in the limit of $N \to
\infty$. Actually, in the limit of $\gamma \ll 1$, the term on the right of Eq.\,(\ref{ws}) which is 
$\propto d_1^2$ is much smaller than the other terms, since $d_1 \ll d_0$.

Before we proceed to the time evolution of the pair-correlation function, we present some numerical results
on $n^{(2)}(\theta, \theta')$. In these results we choose a sufficiently large set of single-particle states,
in order to achieve convergence. In Fig.\,4 we plot the result of such a calculation for $n^{(2)}(\theta, 
\theta'=0)$, for $N = 3, 6$ and 9 atoms, with $\gamma = N g/(2 \pi)$ kept constant and equal to 3. In the limit 
$N \to \infty$, with $N g$ fixed, as we argued earlier, $n^{(2)}(\theta, \theta'=0)$ approaches the horizontal 
line $(N-1)/N \to 1$. 

From Eq.\,(\ref{mfrep}) it is clear that in the limit of large $N$, with $\gamma$ fixed, the pair-correlation
function $n^{(2)}(\theta, \theta')$ will tend to the horizontal line $(N-1)/N$. On the other hand, in the same
limit and for attractive interactions, from Eq.\,(\ref{mfatt}) it follows that $n^{(2)}(\theta, \theta')$ will
be spatially-dependent. 

\begin{figure}[t]
\includegraphics[width=9cm,height=4.5cm,angle=-0]{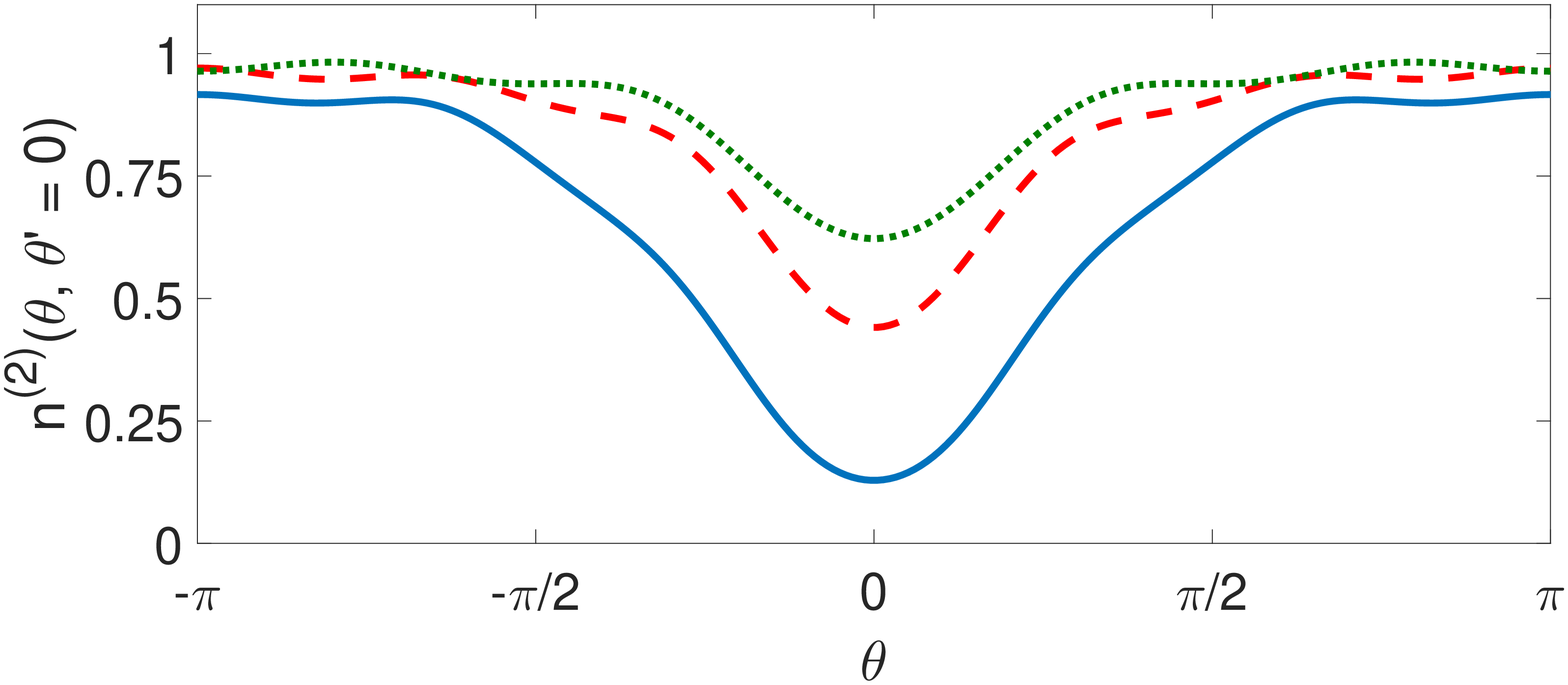}
\caption{(Color online) The pair-correlation function $n^{(2)}(\theta, \theta'=0)$, evaluated numerically, 
for $N = 3$ (blue, solid curve), 6 (red, dashed curve) and 9 (green, dotted curve), with $\gamma = 3$.} 
\end{figure}

\subsection{Time-dependent problem -- general approach}

Turning now to the crucial question of the dynamics, we examine the time-dependent pair-correlation function. 
Without loss of generality, we set $\theta'$ and $t'$ equal to zero, thus considering 
\begin{eqnarray} 
n^{(2)}(\theta, t; \theta'=0, t'=0) = 
  \nonumber \\
  = \frac 
  {\langle \Psi_1 | \Phi^{\dagger}(0, 0) e^{-i H t} \Phi^{\dagger}(\theta, 0) \Phi(\theta, 0) 
  e^{i H t} \Phi(0, 0) | \Psi_1 \rangle} 
  {\langle \Psi_1 | \Phi^{\dagger}(\theta, t)  \Phi(\theta, t) | \Psi_1 \rangle 
   \langle \Psi_1 | \Phi^{\dagger}(0, 0) \Phi(0, 0)| \Psi_1 \rangle}, 
  \label{equivv}
\end{eqnarray}
where $\Phi(\theta, t) = e^{-i H t} \Phi(\theta, 0) e^{i H t}$.

Because of the two hierarchies that we explained earlier, in evaluating the numerator of Eq.\,(\ref{equivv}), 
one may restrict himself to the term $(c_0^{\dagger})^2 c_0^2$, which is of order $N^2$, and to $e^{-i \theta} 
c_0^{\dagger} e^{-i H t} c_0^{\dagger} c_{-1} e^{i H t} c_{1}$ (plus Hermitian conjugate), as well as 
$e^{-i \theta} c_{-1}^{\dagger} e^{-i H t} c_{1}^{\dagger} c_0 e^{i H t} c_{0}$ (plus Hermitian conjugate), 
which are both of order $N d_1$. We stress that the other terms which involve $c_0, c_0^{\dagger}, c_{\pm 1}$ 
and $c_{\pm 1}^{\dagger}$ are of order $N d_1^2$.

In order to evaluate Eq.\,(\ref{equivv}) it is convenient to introduce the two states 
\begin{equation}
| 1 \rangle = c_0 e^{i H t} c_0 |\Psi_1(N,L=0) \rangle,  
\end{equation}
and
\begin{equation}
| 2 \rangle = c_{-1} e^{i H t} c_{1} |\Psi_1(N,L=0) \rangle.
\end{equation}
Starting with the first one, this may be expressed as
\begin{eqnarray}
 |1 \rangle = \sum_{n,m} A_{1,n} B_{n,m} |\Psi_m (N-2, L=0) \rangle,
\end{eqnarray}
where    
\begin{equation}
 A_{1,n} = \langle \Psi_n (N-1, L=0) | c_{0} | \Psi_1 (N, L=0) \rangle e^{i {\cal E}_n(N-1, L=0) t}, 
\end{equation}
and
\begin{equation}
 B_{n,m} = \langle \Psi_m (N-2, L=0) | c_{0} | \Psi_n (N-1, L=0) \rangle. 
\end{equation}
Similarly, 
\begin{eqnarray}
 |2 \rangle = 
 \sum_{n,m} C_{1,n} D_{n,m} |\Psi_m (N-2, L=0) \rangle,
\end{eqnarray}
where 
\begin{eqnarray}
 C_{1,n} =  \langle \Psi_n (N-1, L=-1) | c_{1} | \Psi_1 (N, L=0) \rangle \times
 \nonumber \\
 \times e^{i {\cal E}_n(N-1, L=-1) t}, 
\end{eqnarray}
and
\begin{eqnarray}
 D_{n,m} = \langle \Psi_m (N-2, L=0) | c_{-1} | \Psi_n (N-1, L=-1) \rangle. 
\nonumber \\
\end{eqnarray}
Then, neglecting terms of order $d_1^2/N$, which we examine below, Eq.\,(\ref{equivv}) may be 
written as
\begin{eqnarray} 
n^{(2)}(\theta, t; \theta'=0, t'=0) \approx
\nonumber \\ 
 \approx \frac 1 {N^2} [ A B B^{\dagger} A^{\dagger} + 2 \cos \theta (C D B^{\dagger} A^{\dagger} 
 + A B D^{\dagger} C^{\dagger})] .
\label{finers}
 \end{eqnarray}
This is the final expression for $n^{(2)}(\theta, t; \theta'=0, t'=0)$. The first term on the right is 
of order unity, while the rest are of order $d_1/N$. We have used Eq.\,(\ref{finers}) to evaluate 
numerically the time evolution of the pair-pair correlation function. These results are shown in 
Figs.\,5 and 6. 

\subsection{Time-dependent problem -- approximate, analytic approach}

In addition, because of the assumption of weak interactions, we present below approximate, analytic, 
expressions for the matrices $A$, $B$, $C$, and $D$, and derive a very simple, analytic formula for 
$n^{(2)}(\theta, t; \theta'=0, t'=0)$. 

One crucial observation in the estimates that are presented below is that, because of the assumption of weak 
interactions, the many-body states with a different atom number and/or a different angular momentum are simply 
connected by single-particle excitations. This is also verified from the results that we present below, of the 
numerical diagonalization that we have performed. 

For example, if $|\Psi_1 (N, L=0) \rangle$ is given by Eq.\,(\ref{2st}), then, for weak interactions and large 
$N$,
\begin{eqnarray}
  |\Psi_1 (N-1, L=0) \rangle \propto c_0 |\Psi_1 (N, L=0) \rangle,
\end{eqnarray} 
and therefore
\begin{eqnarray} 
 |\Psi_1 (N-1, L=0) \rangle \approx d_0  |(-1)^0, 0^{N-1}, (+1)^0 \rangle 
\nonumber \\
- d_1 \sqrt{\frac {N-2} {N}} |(-1)^1, 0^{N-3}, (+1)^1 \rangle.
\label{qq1}
\end{eqnarray}
In addition, $|\Psi_2 (N-1, L=0) \rangle$ has to be orthogonal to the above state,
\begin{eqnarray}
  |\Psi_2 (N-1, L=0) \rangle \approx
d_1 \sqrt{\frac {N-2} {N}} |(-1)^0, 0^{N-1}, (+1)^0 \rangle 
\nonumber \\ 
+ d_0 |(-1)^1, 0^{N-3}, (+1)^1 \rangle.
\nonumber \\
\label{qq2}
\end{eqnarray}
Furthermore,
\begin{eqnarray}
  c_0 |\Psi_1 (N, L=0) \rangle \approx \sqrt{N} \, d_0 |(-1)^0, 0^{N-1}, (+1)^0 \rangle 
  \nonumber \\
  - d_1 \sqrt{N-2} \, |(-1)^1, 0^{N-3}, (+1)^1 \rangle 
\end{eqnarray}
and therefore 
\begin{equation}
A \approx
\begin{bmatrix}
 \sqrt{N} e^{i {\cal E}_1(N-1,L=0) t} & {\cal O}(\delta \sqrt{N} e^{i {\cal E}_2(N-1,L=0) t})
\end{bmatrix}.
\end{equation}
About the parameter $\delta$, this is a (very) small quantity, which we define and estimate 
in the following way. If
\begin{eqnarray} 
 |\Psi_1 (N, L=0) \rangle \approx d_0 |(-1)^0, 0^{N}, (+1)^0 \rangle 
 \nonumber \\
 - d_1 |(-1)^1, 0^{N-2}, (+1)^1 \rangle 
 + d_2 |(-1)^2, 0^{N-4}, (+1)^2 \rangle,
\label{eqq1}
\end{eqnarray}
then
\begin{eqnarray} 
 |\Psi_2 (N, L=0) \rangle \approx {\tilde d}_1 |(-1)^0, 0^{N}, (+1)^0 \rangle 
 \nonumber \\
 + {\tilde d}_0 |(-1)^1, 0^{N-2}, (+1)^1 \rangle 
 + {\cal O} ({\tilde d}_1) |(-1)^2, 0^{N-4}, (+1)^2 \rangle,
\nonumber \\
\label{eqq2}
\end{eqnarray}
where $d_0 \approx {\tilde d}_0 \approx 1$ and $d_1 \approx {\tilde d}_1$. Since the two states are 
orthogonal, to leading order, $d_0 {\tilde d}_1 - {\tilde d}_0 d_1 + {\cal O}(d_1 d_2) = 0$. Defining 
$\delta = d_0 {\tilde d}_1 - {\tilde d}_0 d_1$ we conclude that $\delta$ is on the order of $d_1 d_2 
\sim d_1^3$. 

\begin{figure}[t]
\includegraphics[width=9cm,height=4.5cm,angle=-0]{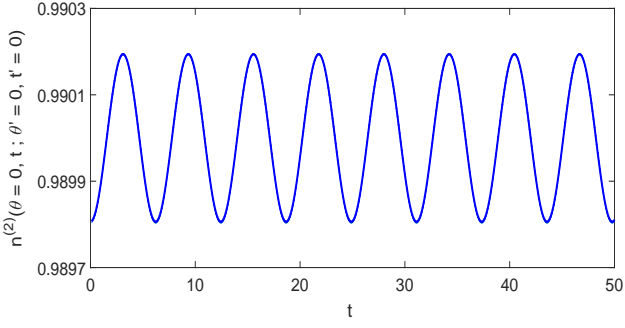}
\caption{(Color online) The time-dependent pair-correlation function $n^{(2)}(\theta = 0, t; \theta'=0, t'=0)$, 
evaluated numerically from Eq.\,(\ref{finers}), for $N = 100$ and $\gamma = 0.01$. The difference between this 
result and the analytic one, Eq.\,(\ref{pairfinnl}), is not visible.} 
\end{figure}

\begin{figure}[t]
\includegraphics[width=9cm,height=4.5cm,angle=-0]{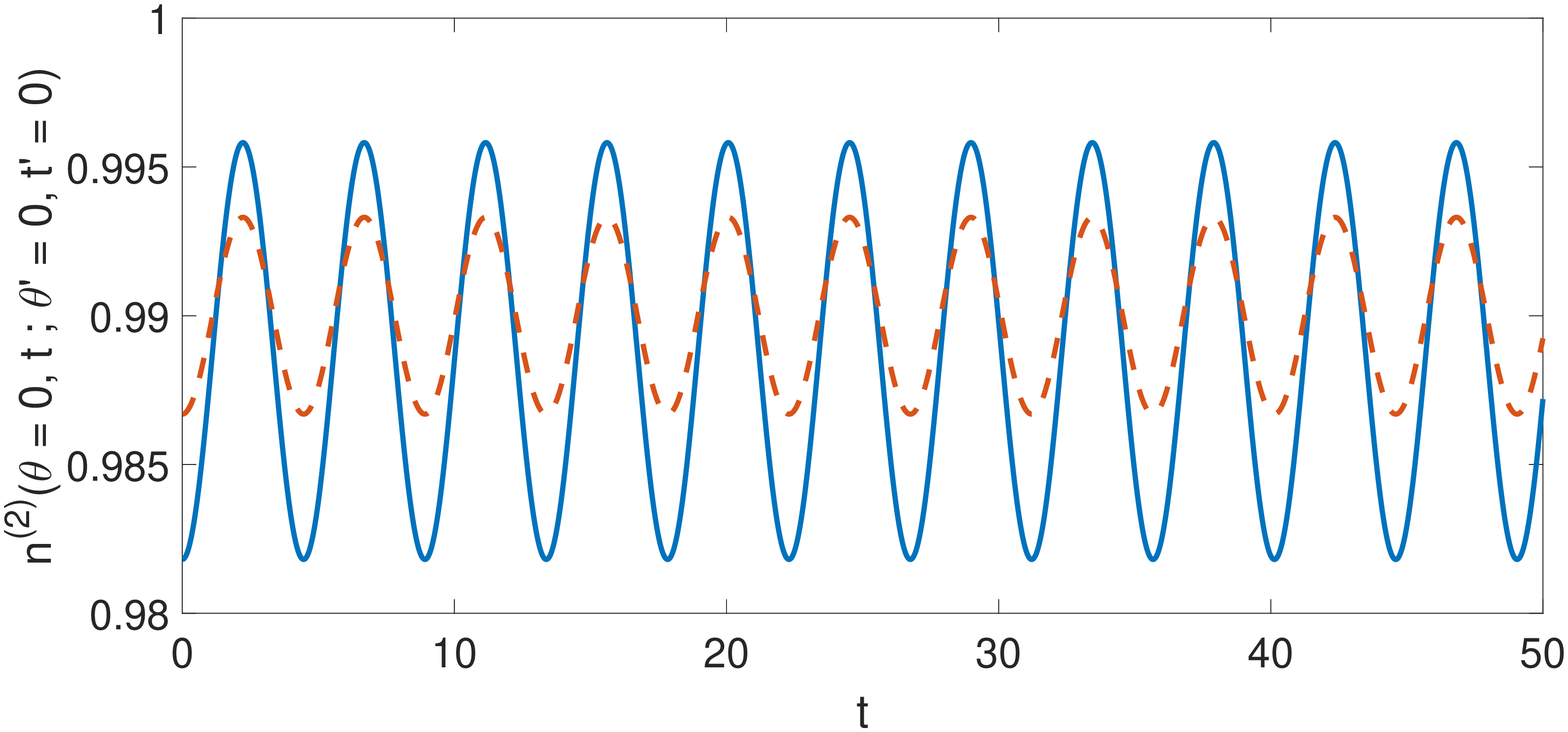}
\caption{(Color online) The time-dependent pair-correlation function $n^{(2)}(\theta = 0, t; \theta'=0, t'=0)$, 
evaluated numerically from Eq.\,(\ref{finers}) (solid, blue curve) and the analytic one, Eq.\,(\ref{pairfinnl}) 
(dashed, red curve), for $N = 100$ and $\gamma = 0.5$.} 
\end{figure}

The above results are in full agreement with those which follow from the numerical diagonalization of the
many-body Hamiltonian. For example, for $N = 100$ atoms, $L = 0$, and $\gamma = 0.05$, showing only the Fock 
states with the three largest amplitudes, the ground state is 
\begin{eqnarray}
 \Psi_1 (N=100, L=0) \rangle \approx 0.9997 |(-1)^0, 0^{100}, (+1)^0 \rangle + 
 \nonumber \\                                       
 - 0.0237 |(-1)^1, 0^{98}, (+1)^1 \rangle + 0.0006 |(-1)^2, 0^{96}, (+1)^2 \rangle,
\nonumber \\
\end{eqnarray}
while the first excited state is 
\begin{eqnarray}
  \Psi_2 (N=100, L=0) \rangle \approx 0.0237 |(-1)^0, 0^{100}, (+1)^0 \rangle +
  \nonumber \\
   + 0.9986 |(-1)^1, 0^{98}, (+1)^1 \rangle -0.0465 |(-1)^2, 0^{96}, (+1)^2 \rangle.
   \nonumber \\
\end{eqnarray}
For the specific choice of parameters $\delta \approx 2.5 \times 10^{-5}$, which is of order $d_1^3$, indeed.
Then, for $N = 99$ atoms, the ground state is
\begin{eqnarray}
|\Psi_1 (N=99, L=0) \rangle \approx 0.9997 |(-1)^0, 0^{99}, (+1)^0 \rangle +
\nonumber \\
 -0.0235 |(-1)^1, 0^{97}, (+1)^1 \rangle + 0.0005 |(-1)^2, 0^{95}, (+1)^2 \rangle,
 \nonumber \\
\end{eqnarray}
while the first excited state is 
\begin{eqnarray}
  \Psi_2 (N=99, L=0) \rangle \approx 0.0235 |(-1)^0, 0^{99}, (+1)^0 \rangle +
  \nonumber \\
  + 0.9987 |(-1)^1, 0^{97}, (+1)^1 \rangle -0.0460 |(-1)^2, 0^{95}, (+1)^2 \rangle.
  \nonumber \\
\end{eqnarray}  

Returning to the matrix $A$,
\begin{equation}
A \approx
\begin{bmatrix}
 \sqrt{N} e^{i {\cal E}_1(N-1,L=0) t} & 0
\end{bmatrix}.
\label{ma}
\end{equation}
We find in a similar way that
\begin{equation}
B \approx
\begin{bmatrix}
\sqrt{N-1} & {\cal O}(d_1^3 \sqrt{N}) \\
{\cal O}(d_1/\sqrt{N}) & \sqrt{N-3} 
\end{bmatrix}
\approx 
\begin{bmatrix}
\sqrt{N-1} & 0 \\
0 & \sqrt{N-3} 
\end{bmatrix}.
\label{mb}
\end{equation}
Here, the off-diagonal matrix elements have a different order of magnitude because of the
different dependence of the amplitudes on $N$ [see, e.g., Eqs.\,(\ref{qq1}) and (\ref{qq2})]. 
Also, since $|\Psi_1 (N-1, L=-1) \rangle \propto c_1 |\Psi_1 (N, L=0) \rangle$, 
it follows directly that
\begin{eqnarray} 
 |\Psi_1 (N-1, L=-1) \rangle \approx - |(-1)^1, 0^{N-2}, (+1)^0 \rangle 
\nonumber \\
+ {\sqrt 2} (d_2/d_1) |(-1)^2, 0^{N-4}, (+1)^1 \rangle 
\nonumber \\
+ {\cal O} (d_3/d_1) |(-1)^3, 0^{N-6}, (+1)^1 \rangle,
\label{qqq1}
\end{eqnarray}
while
\begin{eqnarray}
|\Psi_2 (N-1, L=-1) \rangle \approx {\sqrt 2} (d_2/d_1) |(-1)^1, 0^{N-2}, (+1)^0 \rangle 
\nonumber \\
+ |(-1)^2, 0^{N-4}, (+1)^1 \rangle + {\cal O} (d_2/d_1) |(-1)^3, 0^{N-6}, (+1)^1 \rangle.
\nonumber \\
\end{eqnarray}
Again, the above results are in full agreement with the ones from the diagonalization. For 
example, for $N = 99$ atoms, $L = -1$ and $\gamma = 0.05$, the ground state is
\begin{eqnarray}
|\Psi_1 (N=99, L=-1) \rangle \approx -0.9995 |(-1)^1, 0^{98}, (+1)^0 \rangle +
 \nonumber \\
+0.0329 |(-1)^2, 0^{96}, (+1)^1 \rangle - 0.0009 |(-1)^3, 0^{94}, (+1)^2 \rangle,
\nonumber \\
\end{eqnarray} 
while the first excited state is 
\begin{eqnarray}
  |\Psi_2 (N=99, L=-1) \rangle \approx 0.0329 |(-1)^1, 0^{98}, (+1)^0 \rangle +
  \nonumber \\
  + 0.9979 |(-1)^2, 0^{96}, (+1)^1 \rangle -0.0558 |(-1)^3, 0^{94}, (+1)^2 \rangle.
  \nonumber \\
\end{eqnarray} 

Therefore,
\begin{equation}
C \approx
\begin{bmatrix}
 -d_1 e^{i {\cal E}_1(N-1,L=-1) t} & {\cal O}((d_2 d_3/d_1) e^{i {\cal E}_2(N-1,L=-1) t})
\end{bmatrix}
\nonumber
\end{equation}
\begin{equation}
\approx 
\begin{bmatrix}
 -d_1 e^{i {\cal E}_1(N-1,L=-1) t} & 0 
\end{bmatrix},
\label{mc}
\end{equation}
since $d_2 d_3/d_1$ is ${\cal O} (d_1^4)$. Finally, since $|\Psi_1 (N-2, L=0) \rangle \propto c_{-1} 
|\Psi_1 (N-1, L=-1) \rangle$, 
\begin{equation}
D \approx
\begin{bmatrix}
1 & -d_1 \\
{\cal O}(d_1^3) & \sqrt{2} 
\end{bmatrix}
\approx
\begin{bmatrix}
1 & -d_1 \\
0 & \sqrt{2} 
\end{bmatrix}.
\label{md}
\end{equation}
From the above approximate expressions for the matrices $A$, $B$, $C$ and $D$, Eqs.\,(\ref{ma}), (\ref{mb}), 
(\ref{mc}), and (\ref{md}), Eq.\,(\ref{finers}) implies that 
\begin{eqnarray} 
n^{(2)}(\theta, t; \theta'=0, t'=0) \approx
 \nonumber \\
 \approx \frac {N-1} {N} - 4 d_1 \frac {\sqrt{N (N-1)}} {N^2} 
 \cos \theta \cos (\Delta E \, t),
\label{pairfinnl}
\end{eqnarray}
where 
$\Delta E = {\cal E}_1(N-1, L=-1) - {\cal E}_1(N-1, L=0)$. From Eq.\,(\ref{pairfinnl}) we see that in the limit
of large $N$, with $\gamma$ fixed, $n^{(2)}(\theta, t; \theta'=0, t'=0)$ tends to unity and thus we do not have a
time crystal in this limit (as discussed also below).

For weak interactions, the dominant term in the many-body state with $N-1$ atoms and $L=0$ is $|(-1)^0, 0^{N-1}, 
(+1)^0 \rangle$, while in the state with $N-1$ atoms and $L=-1$ it is $|(-1)^1, 0^{N-2}, (+1)^0 \rangle$. As a 
result, 
\begin{equation}
{\cal E}_1(N-1, L=0) \approx g (N^2 - 3 N + 2)/(4 \pi), 
\end{equation}
while 
\begin{equation}
{\cal E}_1(N-1, L=-1) \approx \epsilon_{-1} + g (N^2 - N - 2)/(4 \pi), 
\end{equation}
and therefore 
\begin{equation}
\Delta E \approx \epsilon_{-1} + g (N-2)/(2 \pi) \approx \epsilon_{-1} + \gamma, 
\end{equation}
(with $\epsilon_{-1} = \epsilon_{1}$). This energy difference determines the period of the 
pair-correlation function $n^{(2)}(\theta, t; \theta'=0, t'=0)$, which is equal to $2 \pi /\Delta E$. 
We stress that the approximate expression of Eq.\,(\ref{pairfinnl}) becomes asymptotically exact for 
large values of $N$ and small values of $\gamma$. 

As mentioned earlier, in evaluating $n^{(2)}(\theta, t; \theta'=0, t'=0)$ from the approximate expression 
of Eq.\,(\ref{finers}), we kept terms up to order $d_1/N$, and neglected terms, which are, to leading order, 
$d_1^2/N$. Interestingly enough, this leading-order correction is $\propto (d_1^2/N) \cos \theta \cos 
(\Delta E \, t)$ and therefore this term has the same spatial and temporal behaviour as the one in 
Eq.\,(\ref{pairfinnl}). As a result, the temporal and spatial dependence of Eq.\,(\ref{finers}) is 
correct to order $d_1^2/N$. 
 
The analytic results presented above are in full agreement with the extended numerical simulations that 
we have performed. Keeping the terms with $m = 0, 1, 2$, and 3 in the many-body state of Eq.\,(\ref{fock}), 
we have evaluated numerically the expression of Eq.\,(\ref{finers}). The result of this calculation 
is shown in Figs.\, 5 and 6, for $N = 100$ and $\gamma = 0.01$ and $\gamma = 0.5$, respectively. For the 
smaller value of $\gamma$, the numerical result, Eq.\,(\ref{finers}) converges to the analytic one (for 
large $N$ and small $\gamma$), i.e., Eq.\,(\ref{pairfinnl}). The difference between the two curves is 
not visible in Fig.\,5. 

On the other hand, for $\gamma = 0.5$, there is a substantial difference between the two curves, as seen 
in Fig.\,6, however this difference is observed in the amplitude only and in the temporal period. This is 
due to two facts. Firstly, the result of Eq.\,(\ref{finers}) is correct to order $d_1^2/N$, as mentioned 
in the previous paragraph. Secondly, the low-lying excited states of the many-body Hamiltonian are 
equidistant \cite{Ueda1, GK, Ueda2}, apart from corrections which are of order $1/N$ \cite{remark}. For 
example, again from the diagonalization of the Hamiltonian, we find that the five lowest eigenenergies for 
$N = 100$, $L = 0$, and $\gamma = 0.5$ are ${\cal E}_1(N=100, L = 0) \approx 2.4738, 
{\cal E}_2(N=100, L = 0) \approx 4.5701, {\cal E}_3(N=100, L = 0) \approx 6.6637, 
{\cal E}_4(N=100, L = 0) \approx 8.7546$, and ${\cal E}_5(N=100, L = 0) \approx 10.8427$, with an
energy difference between successive energy levels which is roughly 2.09.

It is worth commenting on the case of attractive interactions \cite{FW1, Sachaprl, Sacha2}, where it
was shown that for a finite number of atoms there is no time-crystalline behaviour \cite{Sacha2}. Indeed,
for a finite $N$, in the evaluation of the pair-correlation function, there will be mixing of various 
states, according to the results of Sec.\,III B and of Fig.\,2, which will result in a non-crystalline
behaviour of $n^{(2)}(\theta, t; \theta'; t')$. 

On the other hand, in the limit of large atom numbers, the many-body state reduces to a mean-field state, 
with an inhomogeneous density distribution (as opposed to the repulsive interactions, where there is no
spatial dependence of the pair-correlation function). An additional requirement in order to have a time 
crystal is to have rotation, however this is accomplished by exciting the center of mass motion of the 
cloud. For example, for $L = N$ it follows from Eq.\,(\ref{mfatt}) that 
\begin{eqnarray}
\Psi_{\rm MF}(\theta_1, \theta_2, \dots, \theta_N) = 
\nonumber \\
 = e^{i \sum_{i=1}^N \theta_i} 
 \prod_{i=1}^N [a_{-1} \psi_{-1}(\theta_i) + a_{0} \psi_{0}(\theta_i) + a_{1} \psi_{1}(\theta_i)].
\label{mfattt}
\end{eqnarray}
As a result, we do have a time crystal in this limit of large $N$, with $\gamma$ kept fixed
\cite{FW1, Sachaprl}. This case is known as a ``symmetry protected time crystal" because 
it relies on angular-momentum conservation, in an excited state of the system.

\section{Summary, experimental relevance, and conclusions}

Summarizing our results, in the present study we have considered Bose-Einstein condensed atoms, which 
are confined in a very tight annular/toroidal trap, thus realizing a (quasi)-one-dimensional system, 
under periodic boundary conditions. While axial symmetry forces the single-particle density distribution 
to be axially symmetric, the pair-correlation function -- which examines the correlations between the 
atoms -- breaks the axial symmetry of the Hamiltonian, even in the ground state of the system. Furthermore, 
these correlations show a temporal periodic behaviour.

One important feature of the many-body problem that we have considered is that, for repulsive interactions, 
the amplitudes of the Fock states that constitute the many-body state of the gas decay exponentially. 
Actually, this is intimately connected with the fact that we have a Bose-Einstein condensed gas, i.e., 
a single-particle state that is occupied by a macroscopic number of atoms. Clearly this is even more 
pronounced in the limit of weak interactions that we have considered here, where the depletion of the 
condensate is suppressed. On the other hand, the case of sufficiently strong attractive interactions
is very different, where the condensate is fragmented.

The temporal period of the pair-pair correlation function that we have evaluated is set by the energy 
$\epsilon_1 + \gamma$. Since, in our units, $\epsilon_1 = 1$ and $\gamma \ll 1$, the corresponding 
temporal period is set by the energy $\epsilon_1$ of the lowest mode associated with the rotation of 
the atoms around the ring potential. For a typical radius of the ring equal to 20 $\mu$m, $\epsilon_1$ 
is on the order of a few Hertz, however tuning the radius of the ring may change this number significantly. 
For large large $N$ and small $\gamma$, $d_1 \approx \gamma/2$ and the amplitude of $n^{(2)}(\theta, t; 
\theta', t')$ is $8 d_1/N \approx 2 g/\pi = 4 \gamma/N$. Therefore, a choice of, e.g., $N = 1000$ and 
$\gamma = 0.5$ gives an amplitude of 0.002.

For the case of repulsive interactions, the problem that we have studied provides an explicit example where, 
although one gets a periodic temporal behaviour (of the derived pair-correlation function), this quantity 
tends to a constant in the thermodynamic limit of a large atom number, i.e., in the limit $N \to \infty$, 
with $N g \propto \gamma$ fixed. This fact implies that this is not a genuine time crystal. This is in 
agreement with the results of Refs.\,\cite{Watanabe}, where it was shown that in this limit, for correlation 
functions that are sufficiently nonlocal, time crystals are not possible in the ground state of a system. 

The analysis that we have presented also provides insight into the problem of attractive interactions and 
for a rotating system, where, as shown in Refs.\,\cite{FW1, Sachaprl} it is possible to realize a time 
crystal in the limit of a large atom number, with $\gamma$ kept fixed. Still, for a finite $N$ this property 
is lost, as a result of the nature of the many-body state.

The importance of the present study is three fold. Firstly, it analyses the specific quantum system, 
in the case of a small atom number -- away from the thermodynamic limit -- and provides insight into 
the many-body state, including both the ground state, as well as the elementary excitations. Furthermore,
it compares the case of repulsive with the very different case of attractive interactions. Secondly, 
it provides an example where -- for repulsive interactions -- the temporal periodic behaviour of the 
pair-correlation function disappears in the large atom limit, as expected due to very general arguments. 
Thirdly, it sheds light on the general principles underlying the physics of time crystals and, more 
generally, of many-body quantum systems.

\acknowledgements GMK wishes to thank Chris Pethick, Krzysztof Sacha, and Wolf von Klitzing for useful 
discussions.

{\bf Data Availability Statement.} The presented data are available on request from the authors.

{\bf Declarations}

{{\bf Conflict of interest} The authors declare that they have no conflict of interest.}


\begin{thebibliography}{99}
\bibitem{remark} Very generally, from Eq.\,(\ref{equivv}), when $\Phi(0, 0)$ acts
on $|\Psi_1 \rangle$ an atom is taken out of the system. This operation creates a linear superposition 
of states (having various values of the angular momentum, but all of them having $N-1$ atoms). When 
the exponential $e^{i H t}$ acts on this superposition of states, each of these states will have an
exponential $e^{i {\cal E}_n t}$ (here we do not write explicitly the dependence of ${\cal E}_n$ on 
the atom number and on the angular momentum). The same will happen when the second exponential, 
$e^{-i H t}$, acts on $\Phi^{\dagger}(\theta, 0) \Phi(\theta, 0) e^{i H t} \Phi(0, 0) | \Psi_1 
\rangle$, which will give another exponential factor $e^{-i {\cal E}_m t}$. If the energy spectrum 
is equidistant (for all values of the angular momentum, in general), these exponentials will involve
an integer multiple of only one energy and as a result $n^{(2)}(\theta, t; \theta'=0, t'=0)$ will be
a periodic function (in time).
\end{thebibliography}
\end{document}